\documentclass{ptephy_v1}

\preprintnumber{XXXX-XXXX} 

\begin{document}

\title{Charged-current scattering off $^{16}$O nucleus as a detection channel for supernova neutrinos}

\author{Ken'ichiro Nakazato}
\affil{Faculty of Arts \& Science, Kyushu University, 744 Motooka, Nishi-ku, Fukuoka 819-0395, Japan}
\author{Toshio Suzuki}
\affil{Department of Physics, College of Humanities and Sciences, Nihon Univerity, Sakurajosui 3-25-40, Setagaya-ku, Tokyo 156-8550, Japan}
\affil{National Astronomical Observatory of Japan, 2-21-1 Osawa, Mitaka, Tokyo 181-8588, Japan}
\author{Makoto Sakuda}
\affil{Department of Physics, Okayama University, 3-1-1 Tsushimanaka, Kita-ku, Okayama 700-8530, Japan}

\begin{abstract}%
Event spectra of the neutrino-$^{16}$O charged-current reactions in Super-Kamiokande are evaluated for a future supernova neutrino burst. Since these channels are expected to be useful for diagnosing a neutrino spectrum with high average energy, the evaluations are performed not only for an ordinary supernova neutrino model but also for a model of neutrino emission from a black-hole-forming collapse. Using shell model results, whose excitation energies are consistent with the experimental data, the cross sections of the $^{16}$O($\nu_e, e^-$)X and $^{16}$O($\bar\nu_e, e^+$)X reactions for each nuclear state with a different excitation energy are employed in this study. It is found that, owing to the components of the reaction with higher excitation energy, the event spectrum becomes 4--7~MeV softer than that in the case without considering the excitation energies. In addition, a simplified approach to evaluate the event spectra is proposed for convenience and its validity is examined.
\end{abstract}

\subjectindex{E26, F22, D02}

\maketitle

\section{Introduction}
The detection of supernova neutrinos is one of the main targets of terrestrial neutrino experiments~\cite{hirata87,bionta87,alex88}. Core collapse supernovae emit numerous neutrinos of all flavors with energies of tens of MeV, which are expected to provide various information on supernova physics as well as neutrino physics~\cite{kotake12,janka12,burrows13,self13a,fogl15,mirizzi16,janka16,hori18}. Currently, several neutrino detectors are operating and ready for the next Galactic supernova~\cite{mirizzi16,hori18,scholberg12}. Using different kinds of detectors such as the water Cherenkov, liquid scintillator and liquid argon types, we can study the flavor-dependent fluxes and spectra of supernova neutrinos~\cite{laha14a,laha14b,fischer15,lu16,nikrant18,li18,seadrow18}.

We focus on the neutrino-induced reactions in water Cherenkov detectors~\cite{ikeda07}. The dominant channel for supernova neutrino observation is $\bar\nu_e$ absorption on free protons, which is called an inverse $\beta$-decay reaction. Elastic scattering between neutrinos and electrons is also expected. Furthermore, neutrinos interact with $^{16}$O nuclei through charged- and neutral-current channels. In particular, for the charged-current $^{16}$O($\nu_e, e^-$)X and $^{16}$O($\bar\nu_e, e^+$)X reactions, the recoil $e^-$/$e^+$ emits the Cherenkov light, as is the case for the inverse $\beta$-decay reaction and neutrino-electron scattering. Their event number strongly depends on the average energy of neutrinos because the energy thresholds of these interactions are as high as $\sim$15~MeV and $\sim$11~MeV for the $^{16}$O($\nu_e, e^-$)X and $^{16}$O($\bar\nu_e, e^+$)X reactions, respectively~\cite{haxton87}. Therefore, these channels will be useful if the average energy is as high as in the case of neutrinos from a black-hole-forming collapse and the early (accretion) phase of a supernova neutrino burst.

Super-Kamiokande (SK) is the only current example of a large water Cherenkov detector. In June 2018, the collaboration started an upgrade to load gadolinium (Gd) into the water in SK, which is called the SK-Gd project~\cite{sekiya18}. As a result of this update, neutrons produced by the inverse $\beta$-decay reactions are expected to be captured on Gd, leading to the emission of $\gamma$ rays, which are detectable in SK~\cite{bv04}. Thus, taking the coincidence of a prompt positron signal and $\gamma$ rays from the neutron capture (neutron tagging), events of the inverse $\beta$ decay reaction are discriminated from other events. Furthermore, while the cross section of the electron-scattering reaction has a strong forward peak, that of the neutrino-$^{16}$O charged-current reaction has a moderate backward peak~\cite{haxton87}. Accordingly, the neutrino-$^{16}$O charged-current reaction is detected as neither a neutron-tagged nor a forward-scattered event in SK-Gd.

So far, the neutrino-$^{16}$O reactions have been extensively studied for a broad range of the induced neutrino energy~\cite{haxton87,kuramoto90,jachowicz02,kolbe02,kolbe03,bc05,benhar05,lv07,ankow12}. As in the case of supernova neutrinos, neutrinos with energies of tens of MeV can excite $^{16}$O nuclei to giant resonances. Since the energy of the recoil $e^-$/$e^+$ is equal to the difference between the induced neutrino energy and the nuclear excitation energy, the cross section for each excited state in the resonance is required to evaluate the energy distribution of the recoil $e^-$/$e^+$. For the channels of the neutrino-$^{16}$O charged-current reaction, in this study we investigate the event spectra in the SK taking into account the cross section for each excitation energy. In the previous studies~\cite{laha14a,nikrant18,sk16}, the neutrino event numbers are investigated with the cross section including the contributions from individual excited states while the excitation energy of each state is not considered to evaluate the energy of recoil $e^-$/$e^+$. In this study, the neutrino event spectra are investigated using the cross section of each excited state in a consistent manner for the first time.

This paper is organized as follows. In Sect.~\ref{cs}, we present the cross sections of the $^{16}$O($\nu_e, e^-$)X and $^{16}$O($\bar\nu_e, e^+$)X reactions for each nuclear state with different excitation energies. A simplified approach to calculate the energy distribution of the recoil $e^-$/$e^+$ is also proposed. In Sect.~\ref{spct}, we show $e^-$/$e^+$ spectra induced by supernova neutrinos and evaluate the validity of the simplified approach. For this purpose, both analytic expressions and numerical models for the supernova neutrino spectra are adopted. Finally, Sect.~\ref{conc} is devoted to our conclusions.

\section{Cross sections of neutrino-$^{16}$O charged-current reactions}\label{cs}
We first discuss the neutrino-$^{16}$O charged-current reactions. Since the recoil $e^-$/$e^+$ produced by $^{16}$O($\nu_e, e^-$)X and $^{16}$O($\bar\nu_e, e^+$)X reactions is detectable via the Cherenkov light, these reactions are possible detection channels of supernova neutrinos. The total energy (including the rest mass) of the recoil $e^-$/$e^+$, $E_e$, is related to the neutrino energy, $E_\nu$, as $E_e=E_\nu-E_x$, where $E_x$ is the excitation energy of the final nucleus measured from the ground state of $^{16}$O. Therefore, to evaluate the energy distributions of the recoil $e^-$/$e^+$, the cross sections of the $^{16}$O($\nu_e, e^-$)X and $^{16}$O($\bar\nu_e, e^+$)X reactions are needed for each excitation energy. In Ref.~\cite{kolbe02}, the neutrino-$^{16}$O charged-current reaction cross sections were calculated as functions of $E_x$ with a continuum random phase approximation (CRPA) and the resultant total cross sections were shown in table form. While the data in Ref.~\cite{kolbe02} are widely used to evaluate event numbers (e.g., a Monte Carlo simulation of the neutrino burst detection system at SK~\cite{sk16}), the cross section for each excitation energy was not shown.

The cross sections of neutrino-$^{16}$O charged-current reactions presented in this study are based on shell model calculations with the SFO-tls Hamiltonian~\cite{suzuki08,suzuki11}, which is a modified version of the SFO Hamiltonian~\cite{suzuki03}. The SFO can well reproduce both the exclusive and inclusive neutrino-$^{12}$C charged-current reaction cross sections induced by decay-at-rest neutrinos~\cite{suzuki06}. In the SFO-tls Hamiltonian, the $p$-$sd$ cross shell part is improved to properly take into account the tensor interaction~\cite{suzuki08,otsuka10}. Using the multipole expansion of weak hadronic currents, the reaction cross sections induced by $\nu_e$ or $\bar{\nu}_e$ are given as~\cite{walecka75,oconnell72,donnelly76,suzuki06}
\begin{eqnarray}\label{diffcs}
& \displaystyle \left(\frac{d\sigma}{d\Omega}\right)_{\frac{\nu_e}{\bar{\nu}_e}} = & \frac{G_F^2\cos^2\theta_C E_e |\vec{k}|}{4\pi^2} F(Z_f,E_e) \frac{4\pi}{2J_i+1} \nonumber \\
& & \times\, \Biggl\{ \sum_{J=0}^{\infty} \biggl\{ (1+\vec{\nu}\cdot\vec{\beta})\,\bigl|\langle J_f \parallel M_J \parallel J_i\rangle\bigr|^2 \nonumber \\ 
& & \qquad\qquad + \left[1-\hat{\nu}\cdot\vec{\beta}+2(\hat{\nu}\cdot\hat{q})(\hat{q}\cdot\vec{\beta})\right]\,\bigl|\langle J_f\parallel L_J \parallel J_i\rangle\bigr|^2 \nonumber \\
& & \qquad\qquad -\, \hat{q}\cdot(\hat{\nu}+\vec{\beta}) \, 2 {\rm Re}\Bigl[ \langle J_f \parallel L_J \parallel J_i\rangle \langle J_f \parallel M_J \parallel J_i\rangle^{\ast}\Bigr] \biggr\} \nonumber \\
& & \qquad + \sum_{J=1}^{\infty} \left[1-(\hat{\nu}\cdot\hat{q})(\hat{q}\cdot\vec{\beta})\right] \nonumber \\
& & \qquad\qquad\quad \times \Bigl(\bigl|\langle J_f \parallel T_{J}^{\rm el} \parallel J_i \rangle\bigr|^2 + \bigl|\langle J_f \parallel T_{J}^{\rm mag} \parallel J_i\rangle\bigr|^2 \nonumber \\
& & \qquad\qquad\qquad\quad \pm\, \hat{q}\cdot(\hat{\nu}-\vec{\beta}) \, 2 {\rm Re} \Bigl[\langle J_f \parallel T_{J}^{\rm mag} \parallel J_i \rangle \langle J_f \parallel T_{J}^{\rm el}\parallel J_i \rangle^{\ast}\Bigr]\Bigr)\Biggr\},
\end{eqnarray}
where $M_{J}$, $L_{J}$, $T_{J}^{\rm el}$ and $T_{J}^{\rm mag}$ are the Coulomb, longitudinal, transverse electric and transverse magnetic multipole operators for the weak hadronic currents, respectively, which are defined by the sum of the vector and axial-vector currents for the charged-current reactions ($\nu_e$, $e^{-}$) and ($\bar{\nu}_e, e^{+}$). The reduced matrix elements of these operators between the initial state $J_i$ and the final state $J_f$ are involved in the cross sections. In Eq.~(\ref{diffcs}), the weak coupling constant is $G_F\cos\theta_C$, where the Fermi coupling constant is $G_{F}$, the Cabibbo angle is $\theta_C$ and the Coulomb correction is taken into account by the Fermi function $F(Z_f,E_e)$ with the charge of the final nucleus $Z_f$~\cite{wm74}. Meanwhile, $\vec{\nu}$ and $\vec{k}$ are neutrino and lepton momenta, respectively, and the other vector quantities are $\vec{q}=\vec{k}-\vec{\nu}$, $\vec{\beta}=\vec{k}/E_e$, $\hat{\nu}=\vec{\nu}/|\vec{\nu}|$ and $\hat{q}=\vec{q}/|\vec{q}|$.

We select 42 states with different excitation energies, as listed in Tables~\ref{exlist1} and \ref{exlist2}, and evaluate the partial cross sections for each state. Note that in the case of $^{16}$O, the dominant contributions are from the spin-dipole transitions. We therefore take transitions to 0$^{-}$, 1$^{-}$ and 2$^{-}$ states and add some transitions to 1$^{+}$ states and a 3$^{-}$ state. The excitation energies of the low-lying 0$^{-}$, 1$^{-}$ and 2$^{-}$ states in $^{16}$F are in good agreement with the experimental data~\cite{as86,tilley93,suda04,wakasa11}. Since the isospin conservation is fully taken into account in the present shell model calculation, the $^{16}$O($\nu_e, e^-$)X and $^{16}$O($\bar\nu_e, e^+$)X reactions have the same excitation energies measured from the ground state of the final nucleus for the 0$^{-}$, 1$^{-}$, 2$^{-}$ and 3$^{-}$ states. Here the differences in the ground-state energies between $^{16}$O and the final nucleus, 14.91~MeV for $^{16}$O($\nu_e, e^-$)X and 10.93~MeV for $^{16}$O($\bar\nu_e, e^+$)X, are taken from the experimental values. In contrast, for the 1$^{+}$ states, the excitation energies of the first 1$^{+}$ states, 3.76~MeV for $^{16}$O($\nu_e, e^-$)X and 3.35~MeV for $^{16}$O($\bar\nu_e, e^+$)X, are taken from the experimental values.
\begin{table}[t]
\caption{Selected states of $^{16}$F and their excitation energies for $^{16}$O($\nu_e, e^-$)X reaction. The excitation energy, $\varepsilon_x$, is measured from the ground state of the final nucleus $^{16}$F and is related to $E_x$ as $\varepsilon_x=E_x-E_{x,\,{\rm g.s.}}$, where $E_{x,\,{\rm g.s.}}$ is the excitation energy of the reaction to the ground state of the final nucleus and $E_{x,\,{\rm g.s.}}=14.91$~MeV for the $^{16}$O($\nu_e, e^-$)X reaction.}
\label{exlist1}
\centering\small
\begin{tabular}{c|cc|cc|cc|cc|cc}
\hline
& & $\varepsilon_x$ & & $\varepsilon_x$& & $\varepsilon_x$ & & $\varepsilon_x$ & & $\varepsilon_x$ \\ 
group & state & (MeV) & state & (MeV) & state & (MeV) & state & (MeV) & state & (MeV) \\ 
\hline
1 & 0$^-$ & 0.00 & 1$^-$ & 0.25 & 2$^-$ & 0.30 & 3$^-$ & 0.34 & 1$^+$ & 3.76 \\
2 & & & 1$^-$ & 5.66 & 2$^-$ & 4.51 & & & 1$^+$ & 4.52 \\
& & & & 8.37 & & 6.71 & & & & 5.78 \\
& & & &  & & 7.57 & & & & 7.03 \\
& & & &  & & & & & & 8.06 \\
3 & & & 1$^-$& 10.61 & 2$^-$ & 12.35 & & & 1$^+$ & 9.52 \\
& & & & 10.81 & & & & & & 9.95 \\
& & & & 11.82 & & & & & & 11.69 \\
& & & & 12.31 & & & & & & 12.51 \\
4 & 0$^-$ & 12.67 & 1$^-$& 13.24 & 2$^-$ & 12.92 & & & 1$^+$ & 12.90 \\
& & 13.08 & & 13.61 & & 13.20 & & & & 13.18 \\
& & & & 14.29 & & 14.02 & & & & 15.06 \\
& & & & 14.45 & & & & & & 16.18 \\
& & & & 15.07 & & & & & & 16.79 \\
& & & & 15.34 & & & & & & 18.08 \\
& & & & 15.90 & & & & & & \\
& & & & 16.85 & & & & & & \\
\hline
\end{tabular}
\end{table}
\begin{table}[t]
\caption{Selected states of $^{16}$N and their excitation energies for $^{16}$O($\bar\nu_e, e^+$)X reaction, for which $E_{x,\,{\rm g.s.}}=10.93$~MeV.}
\label{exlist2}
\centering\small
\begin{tabular}{c|cc|cc|cc|cc|cc}
\hline
& & $\varepsilon_x$ & & $\varepsilon_x$& & $\varepsilon_x$ & & $\varepsilon_x$ & & $\varepsilon_x$ \\ 
group & state & (MeV) & state & (MeV) & state & (MeV) & state & (MeV) & state & (MeV) \\ 
\hline
1 & 0$^-$ & 0.00 & 1$^-$ & 0.25 & 2$^-$ & 0.30 & 3$^-$ & 0.34 & 1$^+$ & 3.35 \\
2 & & & 1$^-$ & 5.66 & 2$^-$ & 4.51 & & & 1$^+$ & 4.12 \\
& & & & 8.37 & & 6.71 & & & & 5.37 \\
& & & & & & 7.57 & & & & 6.63 \\
& & & & & & & & & & 7.66 \\
3 & & & 1$^-$ & 10.61 & 2$^-$ & 12.35 & & & 1$^+$ & 9.12 \\
& & & & 10.81 & & & & & & 9.55 \\
& & & & 11.82 & & & & & & 11.28 \\
& & & & 12.31 & & & & & & 12.10 \\
& & & & & & & & & & 12.49 \\
4 & 0$^-$ & 12.67 & 1$^-$ & 13.24 & 2$^-$ & 12.92 & & & 1$^+$ & 12.77 \\
& & 13.08 & & 13.61 & & 13.20 & & & & 14.66 \\
& & & & 14.29 & & 14.02 & & & & 15.77 \\
& & & & 14.45 & & & & & & 16.38 \\
& & & & 15.07 & & & & & & 17.67 \\
& & & & 15.34 & & & & & & \\
& & & & 15.90 & & & & & & \\
& & & & 16.85 & & & & & & \\
\hline
\end{tabular}
\end{table}

\begin{figure}[t]
\centering\includegraphics[width=5.9in]{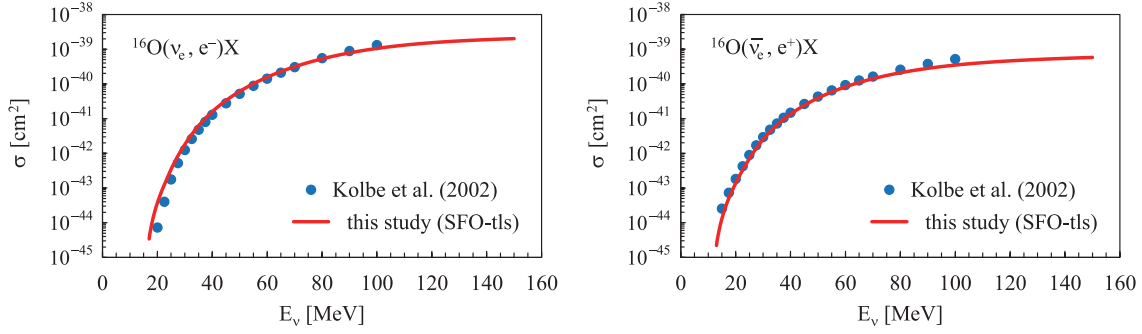}
\caption{Total cross sections of $^{16}$O($\nu_e, e^-$)X reaction (left panel) and $^{16}$O($\bar\nu_e, e^+$)X reaction (right panel). In both panels, the results for the SFO-tls shell model in this study (solid lines) are compared with those for the CRPA model~\cite{kolbe02} (plots).}
\label{totcsoxy}
\end{figure}

\begin{figure}[t]
\centering\includegraphics[width=5.9in]{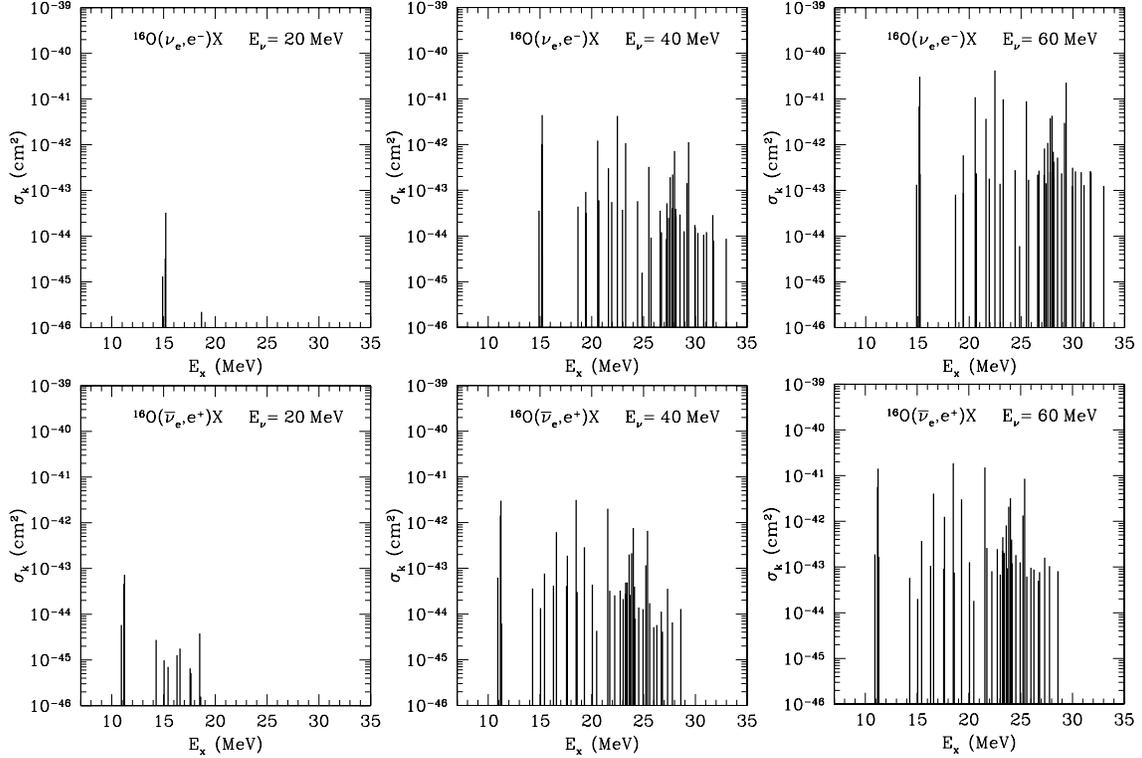}
\caption{The cross sections of $^{16}$O($\nu_e, e^-$)X reaction (upper panels) and $^{16}$O($\bar\nu_e, e^+$)X reaction (lower panels) as functions of the excitation energy. Left, central and right panels are for the neutrino energy $E_\nu=20$, 40 and 60~MeV, respectively.}
\label{partcsoxy}
\end{figure}

The sum of the 42 partial cross sections corresponds to the total cross section, $\sigma(E_\nu)$. In Fig.~\ref{totcsoxy}, we compare the total cross sections of our model (SFO-tls shell model) and the CRPA model~\cite{kolbe02}. In comparison with the CRPA model, the SFO-tls model has a larger cross section for the $^{16}$O($\nu_e, e^-$)X reaction but a smaller cross section for the $^{16}$O($\bar\nu_e, e^+$)X reaction if neutrinos have energies of tens of MeV. Nevertheless, the SFO-tls model and CRPA model have qualitatively similar curves for the total cross sections. It is consistent with the fact that the both models are tested to reproduce the total muon capture rates for $^{16}$O well~\cite{kolbe03,suzuki18}. As a result, it is also expected that the both models have similar strengths of giant resonances. Note that the energy range of $E_\nu \lesssim 100$~MeV is important for supernova neutrino detection, and the contribution of quasi-elastic scattering becomes dominant for $E_\nu \gtrsim 100$~MeV~\cite{benhar05,ankow12}. From Fig.~\ref{partcsoxy}, which shows the cross sections as functions of the excitation energy, we can recognize that the region of the giant resonances is sufficiently included in our model so as to evaluate the spectra of supernova neutrino events. On the other hand, for the neutrino energy above $\sim$100~MeV, our model underestimates the cross section because the contributions of quasi-elastic scattering and states with the higher excitation energies are omitted.

Now we move on to the study of a simplified approach to calculate the energy distribution of the recoil $e^-$/$e^+$. Clearly, we can evaluate the event spectrum, $dN(E_e)/dE_e$, by summing the 42 partial cross sections referred to above as
\begin{equation}\label{full}
\frac{dN(E_e)}{dE_e} = N_T \sum_{k=1}^{42} \left. \sigma_k(E_\nu) \cdot \frac{d{\cal F}(E_\nu)}{dE_\nu} \right|_{E_\nu = E_e+E_{x,k}},
\end{equation}
where $N_T$ is the number of $^{16}$O targets and $d{\cal F}(E_\nu)/dE_\nu$ is the flux of induced neutrinos. The index $k$ denotes the states with excitation energy $E_{x,k}$ and partial cross section $\sigma_k(E_\nu)$. Nevertheless, many of the states have a minor contribution to the total cross section and the excitation energies of some states are very close to each other. Here, we divide the 42 states into four energy groups (Tables~\ref{exlist1} and \ref{exlist2}). Then, the sum of the partial cross sections, $\widetilde\sigma_g(E_\nu)$, is calculated for group $g$ as shown in Table~\ref{cslist}. Note that $\widetilde\sigma_g(E_\nu)$ is related to $\sigma(E_\nu)$ and $\sigma_k(E_\nu)$ as $\sigma(E_\nu)=\sum_{g=1}^{4}\widetilde\sigma_g(E_\nu)=\sum_{k=1}^{42}\sigma_k(E_\nu)$. Subsequently, for each group, we select the representative state that has the largest contribution to the cross section. In our approach, the event spectrum is written as
\begin{equation}\label{approx}
\frac{dN(E_e)}{dE_e} = N_T \sum_{g=1}^{4} \left. \widetilde\sigma_g(E_\nu) \cdot \frac{d{\cal F}(E_\nu)}{dE_\nu} \right|_{E_\nu = E_e+\widetilde E_{x,g}},
\end{equation}
where $\widetilde E_{x,g}$ is the excitation energy of the representative state of group $g$ as shown in Table~\ref{exetc}. The validity of our approach is examined in the next section. Furthermore, for the energy range of $E_\nu<100$~MeV, we fit $\widetilde\sigma_g(E_\nu)$ with the analytic expression
\begin{subequations}\label{csfit}
\begin{equation}\label{csfit0}
\log_{10}\left(\frac{\widetilde\sigma_g(E_\nu)}{\mathrm{cm}^2}\right) \approx a_g+b_g\Lambda(E_\nu)+c_g\left\{\Lambda(E_\nu)\right\}^2,
\end{equation}
\begin{equation}\label{csfit1}
\Lambda(E_\nu)=\log_{10}\left\{\left(\frac{E_\nu}{\mathrm{MeV}}\right)^{1/4}-\left(\frac{\widetilde E_{x,g}}{\mathrm{MeV}}\right)^{1/4}\right\},
\end{equation}
\end{subequations}
which is an extension of the fitting formula in Ref.~\cite{tomas03}. The fitting parameters $a_g$, $b_g$ and $c_g$ are shown in Table~\ref{exetc}.

\begin{table}[t]
\caption{Cross sections of group $g$, $\widetilde\sigma_g(E_\nu)$, for $^{16}$O($\nu_e, e^-$)X and $^{16}$O($\bar\nu_e, e^+$)X reactions as functions of the neutrino energy, $E_\nu$.}
\label{cslist}
\centering\small
\begin{tabular}{r|cccc|cccc}
\hline
\multicolumn{1}{c|}{$E_\nu$} & \multicolumn{4}{c|}{$^{16}$O($\nu_e, e^-$)X} & \multicolumn{4}{c}{$^{16}$O($\bar\nu_e, e^+$)X} \\
\multicolumn{1}{c|}{(MeV)} & $\widetilde\sigma_1$ (cm$^2$) & $\widetilde\sigma_2$ (cm$^2$) & $\widetilde\sigma_3$ (cm$^2$) & $\widetilde\sigma_4$ (cm$^2$) & $\widetilde\sigma_1$ (cm$^2$) & $\widetilde\sigma_2$ (cm$^2$) & $\widetilde\sigma_3$ (cm$^2$) & $\widetilde\sigma_4$ (cm$^2$) \\
\hline
12 & 0.00E$+00$ & 0.00E$+00$ & 0.00E$+00$ & 0.00E$+00$ & 3.03E$-46$ & 0.00E$+00$ & 0.00E$+00$ & 0.00E$+00$ \\
15 & 0.00E$+00$ & 0.00E$+00$ & 0.00E$+00$ & 0.00E$+00$ & 1.32E$-44$ & 0.00E$+00$ & 0.00E$+00$ & 0.00E$+00$ \\
18 & 9.61E$-45$ & 0.00E$+00$ & 0.00E$+00$ & 0.00E$+00$ & 6.16E$-44$ & 9.77E$-46$ & 0.00E$+00$ & 0.00E$+00$ \\
21 & 6.20E$-44$ & 3.24E$-46$ & 0.00E$+00$ & 0.00E$+00$ & 1.73E$-43$ & 2.35E$-44$ & 1.01E$-46$ & 0.00E$+00$ \\
24 & 2.03E$-43$ & 2.78E$-44$ & 0.00E$+00$ & 0.00E$+00$ & 3.81E$-43$ & 1.21E$-43$ & 1.84E$-44$ & 0.00E$+00$ \\
27 & 4.94E$-43$ & 2.22E$-43$ & 1.93E$-45$ & 0.00E$+00$ & 7.21E$-43$ & 3.46E$-43$ & 1.11E$-43$ & 4.56E$-44$ \\
30 & 1.01E$-42$ & 7.31E$-43$ & 1.73E$-44$ & 3.62E$-44$ & 1.23E$-42$ & 7.64E$-43$ & 3.15E$-43$ & 2.09E$-43$ \\
33 & 1.83E$-42$ & 1.73E$-42$ & 6.66E$-44$ & 3.10E$-43$ & 1.94E$-42$ & 1.44E$-42$ & 6.70E$-43$ & 5.32E$-43$ \\
36 & 3.07E$-42$ & 3.41E$-42$ & 1.83E$-43$ & 9.72E$-43$ & 2.88E$-42$ & 2.46E$-42$ & 1.22E$-42$ & 1.06E$-42$ \\
39 & 4.80E$-42$ & 6.00E$-42$ & 4.14E$-43$ & 2.16E$-42$ & 4.08E$-42$ & 3.86E$-42$ & 2.00E$-42$ & 1.83E$-42$ \\
42 & 7.14E$-42$ & 9.74E$-42$ & 8.19E$-43$ & 4.03E$-42$ & 5.56E$-42$ & 5.72E$-42$ & 3.05E$-42$ & 2.90E$-42$ \\
45 & 1.02E$-41$ & 1.49E$-41$ & 1.47E$-42$ & 6.74E$-42$ & 7.32E$-42$ & 8.08E$-42$ & 4.41E$-42$ & 4.32E$-42$ \\
50 & 1.70E$-41$ & 2.70E$-41$ & 3.30E$-42$ & 1.36E$-41$ & 1.09E$-41$ & 1.32E$-41$ & 7.45E$-42$ & 7.55E$-42$ \\
55 & 2.63E$-41$ & 4.45E$-41$ & 6.39E$-42$ & 2.42E$-41$ & 1.52E$-41$ & 1.97E$-41$ & 1.15E$-41$ & 1.20E$-41$ \\
60 & 3.82E$-41$ & 6.78E$-41$ & 1.11E$-41$ & 3.92E$-41$ & 2.03E$-41$ & 2.75E$-41$ & 1.68E$-41$ & 1.76E$-41$ \\
65 & 5.29E$-41$ & 9.71E$-41$ & 1.77E$-41$ & 5.92E$-41$ & 2.58E$-41$ & 3.64E$-41$ & 2.31E$-41$ & 2.45E$-41$ \\
70 & 7.00E$-41$ & 1.32E$-40$ & 2.64E$-41$ & 8.44E$-41$ & 3.18E$-41$ & 4.61E$-41$ & 3.05E$-41$ & 3.25E$-41$ \\
80 & 1.11E$-40$ & 2.16E$-40$ & 5.05E$-41$ & 1.51E$-40$ & 4.41E$-41$ & 6.62E$-41$ & 4.81E$-41$ & 5.10E$-41$ \\
90 & 1.57E$-40$ & 3.09E$-40$ & 8.29E$-41$ & 2.34E$-40$ & 5.60E$-41$ & 8.50E$-41$ & 6.83E$-41$ & 7.10E$-41$ \\
100 & 2.06E$-40$ & 3.98E$-40$ & 1.21E$-40$ & 3.28E$-40$ & 6.68E$-41$ & 1.01E$-40$ & 8.99E$-41$ & 9.06E$-41$ \\
\hline
\end{tabular}
\end{table}

\begin{table}[t]
\caption{Representative excitation energy of group $g$, $\widetilde E_{x,g}$, and fitting parameters in Eq.~(\ref{csfit}) for $^{16}$O($\nu_e, e^-$)X and $^{16}$O($\bar\nu_e, e^+$)X reactions.}
\label{exetc}
\centering
\begin{tabular}{lccccc}
\hline
reaction & group $g$ & $\widetilde E_{x,g}$ (MeV) & $a_g$ & $b_g$ & $c_g$ \\
\hline
$^{16}$O($\nu_e, e^-$)X & 1 & 15.21 & $-40.008$ & 4.918 & 1.036 \\
 & 2 & 22.47 & $-39.305$ & 4.343 & 0.961 \\
 & 3 & 25.51 & $-39.655$ & 5.263 & 1.236 \\
 & 4 & 29.35 & $-39.166$ & 3.947 & 0.901 \\
\hline
$^{16}$O($\bar\nu_e, e^+$)X & 1 & 11.23 & $-40.656$ & 4.528 & 0.887 \\
 & 2 & 18.50 & $-40.026$ & 4.117 & 0.895 \\
 & 3 & 21.54 & $-40.060$ & 3.743 & 0.565 \\
 & 4 & 25.38 & $-39.862$ & 3.636 & 0.846 \\
\hline
\end{tabular}
\end{table}

\section{Spectra of recoil $e^-$/$e^+$ from neutrino-$^{16}$O charged-current reactions}\label{spct}
In this section, we investigate the event spectra induced by the neutrino-$^{16}$O charged-current reactions. In the following, we predict the neutrino events detected in SK with a fiducial volume of 32~kton assuming a supernova at a distance of $d_\mathrm{SN}=10$~kpc and 100\% detection efficiency for $E_e \ge 5$~MeV. Note that in the actual detectors, the excited nuclei produced by the $^{16}$O($\nu_e, e^-$)X and $^{16}$O($\bar\nu_e, e^+$)X reactions may emit extra $\gamma$ rays in their decaying processes, which are obstacles to the accurate event reconstruction. We assume that we can reconstruct the primary $e^-$/$e^+$ energy in the present work and that we can distinguish the extra $\gamma$ rays above 5~MeV from the primary $e^-$/$e^+$ by checking the Cherenkov angle around the primary vertex \cite{sk05,sk12} or Cherenkov-ring counting method \cite{sk17}. To examine the validity of the simplified approach introduced in the previous section, the event spectra evaluated with Eqs.~(\ref{full}) and (\ref{approx}) are compared. Hereafter, we refer to the spectra evaluated with Eqs.~(\ref{full}) and (\ref{approx}) as ``42-state case'' and ``four-group case'', respectively. For the supernova neutrino spectra, we first adopt the analytic expressions in Sect.~\ref{spct1}. As the next step, numerical models are utilized in Sect.~\ref{spct2}.

\subsection{Analytic case for supernova neutrino spectra}\label{spct1}
Here, we consider the time-integrated flux of supernova neutrinos represented as
\begin{equation}\label{snnflux}
\frac{d{\cal F}(E_\nu)}{dE_\nu} = \frac{1}{4\pi d^2_\mathrm{SN}}\frac{E_{\nu_i,\mathrm{tot}}}{\langle E_{\nu_i}\rangle}f(E_\nu),
\end{equation}
where subscript $i$ denotes the species of neutrinos, i.e., $\nu_i=\nu_e$, $\bar\nu_e$. While the total energy emitted by the $\nu_i$ flavor is set to $E_{\nu_i,\mathrm{tot}}=5\times10^{52}$~erg, two values are chosen for the average energy of $\nu_i$, $\langle E_{\nu_i}\rangle=10$~MeV and 20~MeV. For the normalized neutrino distribution function, $f(E_\nu)$, we consider the following three models: (i) Fermi--Dirac distribution, (ii) Maxwell--Boltzmann distribution and (iii) modified Maxwell--Boltzmann distribution. For model (i), we take the spectrum with zero chemical potential written as
\begin{equation}\label{spectfd}
f_\mathrm{FD}(E_\nu) = \frac{2}{3\zeta(3)T_{\nu_i}^3}\frac{E_\nu^2}{\exp(E_\nu/T_{\nu_i})+1}
\end{equation}
with the neutrino temperature $T_{\nu_i}=\frac{180}{7\pi^4}\zeta(3)\langle E_{\nu_i}\rangle\approx\langle E_{\nu_i}\rangle /3.151$ and the zeta function $\zeta(3)\approx 1.202$. On the other hand, the spectral forms are
\begin{equation}\label{spectmb}
f_\mathrm{MB}(E_\nu) = \frac{27}{2\langle E_{\nu_i}\rangle^3}E_\nu^2\exp\left(-\frac{3E_\nu}{\langle E_{\nu_i}\rangle}\right)
\end{equation}
for model (ii) and
\begin{equation}\label{spectmmb}
f_\mathrm{mMB}(E_\nu) = \frac{128}{3\langle E_{\nu_i}\rangle^4}E_\nu^3\exp\left(-\frac{4E_\nu}{\langle E_{\nu_i}\rangle}\right)
\end{equation}
for model (iii). Incidentally, they are generalized as
\begin{equation}\label{spectgmb}
f_\alpha(E_\nu) = \frac{(\alpha+1)^{\alpha+1}}{\Gamma(\alpha+1)\langle E_{\nu_i}\rangle^{\alpha+1}}E_\nu^\alpha\exp\left(-\frac{(\alpha+1)E_\nu}{\langle E_{\nu_i}\rangle}\right)
\end{equation}
with the gamma function $\Gamma(\alpha+1)$ where $\alpha=2$ and 3 correspond to models (ii) and (iii), respectively~\cite{keil03}. Here, $\alpha$ is referred to as the shape parameter. The spectrum is more pinched (high-energy tail suppressed) for larger $\alpha$. For the second energy moment, $\langle E^2_{\nu_i}\rangle$, Eq.~(\ref{spectgmb}) gives $\langle E^2_{\nu_i}\rangle / \langle E_{\nu_i}\rangle^2=(2+\alpha)/(1+\alpha)$. Since Eq.~(\ref{spectfd}) gives $\langle E^2_{\nu_i}\rangle / \langle E_{\nu_i}\rangle^2=1.303$, model (i) corresponds to $\alpha=2.301$. Another way to give pinched spectra is Fermi--Dirac distribution with a nonzero chemical potential, which is proportional to $E_\nu^2/[\exp(E_\nu/T-\eta)+1]$ with parameters $T$ and $\eta$, where $\eta>0$ indicates a pinched spectrum~\cite{ds00}. Based on again the value of $\langle E^2_{\nu_i}\rangle / \langle E_{\nu_i}\rangle^2$, $\alpha=2$ and 3 correspond to $\eta=-\infty$ and 1.694, respectively.

\begin{figure}[t]
\centering\includegraphics[width=5.9in]{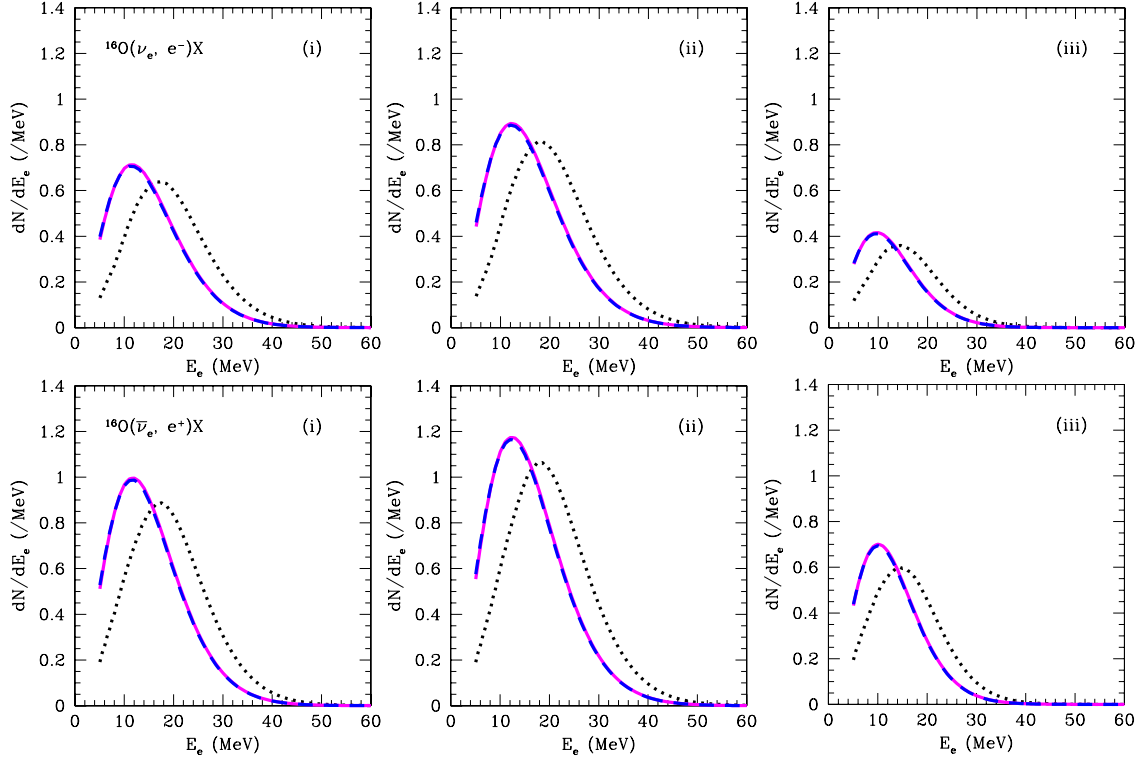}
\caption{Event spectra for $^{16}$O($\nu_e, e^-$)X reaction (upper panels) and $^{16}$O($\bar\nu_e, e^+$)X reaction (lower panels) under the assumptions of the (i) Fermi-Dirac distribution [Eq.~(\ref{spectfd})], (ii) Maxwell-Boltzmann distribution [Eq.~(\ref{spectmb})] and (iii) modified Maxwell-Boltzmann distribution [Eq.~(\ref{spectmmb})] for the supernova neutrino spectra. The neutrino average energy is taken to be $\langle E_{\nu_e}\rangle=\langle E_{\bar\nu_e}\rangle=10$~MeV. Solid, dashed and dotted lines represent ``42-state case'', ``four-group case'' and ``single-energy case'', respectively.}
\label{ana10}
\end{figure}

In Fig.~\ref{ana10}, we show the event spectra of the recoil $e^-/e^+$ produced by the $^{16}$O($\nu_e, e^-$)X and $^{16}$O($\bar\nu_e, e^+$)X reactions for $\langle E_{\nu_e}\rangle=\langle E_{\bar\nu_e}\rangle=10$~MeV. We can see that the spectrum of the ``four-group case'', which is the simplified approach introduced in this study, is very similar to that of the more accurate ``42-state case'' for each condition. Here, we also show the spectra calculated under the assumption that all of the recoil $e^-$/$e^+$ have an energy of $E_e = E_\nu-E_{x,\,{\rm g.s.}}$, where $E_{x,\,{\rm g.s.}}$ is the excitation energy of the reaction to the ground state of the final nucleus, which we refer to as the ``single-energy case''. Note that this type of estimation has usually been carried out so far (e.g., \cite{nikrant18}). The ``single-energy case'' has a harder spectrum than the ``42-state case'' and ``four-group case'' because the estimation of $E_e$ is about 4~MeV too high in the ``single-energy case''. In other words, a considerable fraction of neutrino-$^{16}$O charged-current reactions occur with an excitation energy clearly higher than the threshold energy. Regarding the difference in the induced supernova neutrino spectrum, a model (ii) has a larger event number than model (iii). This is because, for a fixed neutrino average energy, the Maxwell--Boltzmann distribution has a harder spectrum and is more abundant in high-energy neutrinos than the modified Maxwell--Boltzmann distribution. Since the hardness of the Fermi--Dirac distribution is between those of the Maxwell--Boltzmann and modified Maxwell--Boltzmann distributions, the event number of model (i) is larger than that of model (iii) but smaller than that of model (ii).

\begin{figure}[t]
\centering\includegraphics[width=5.9in]{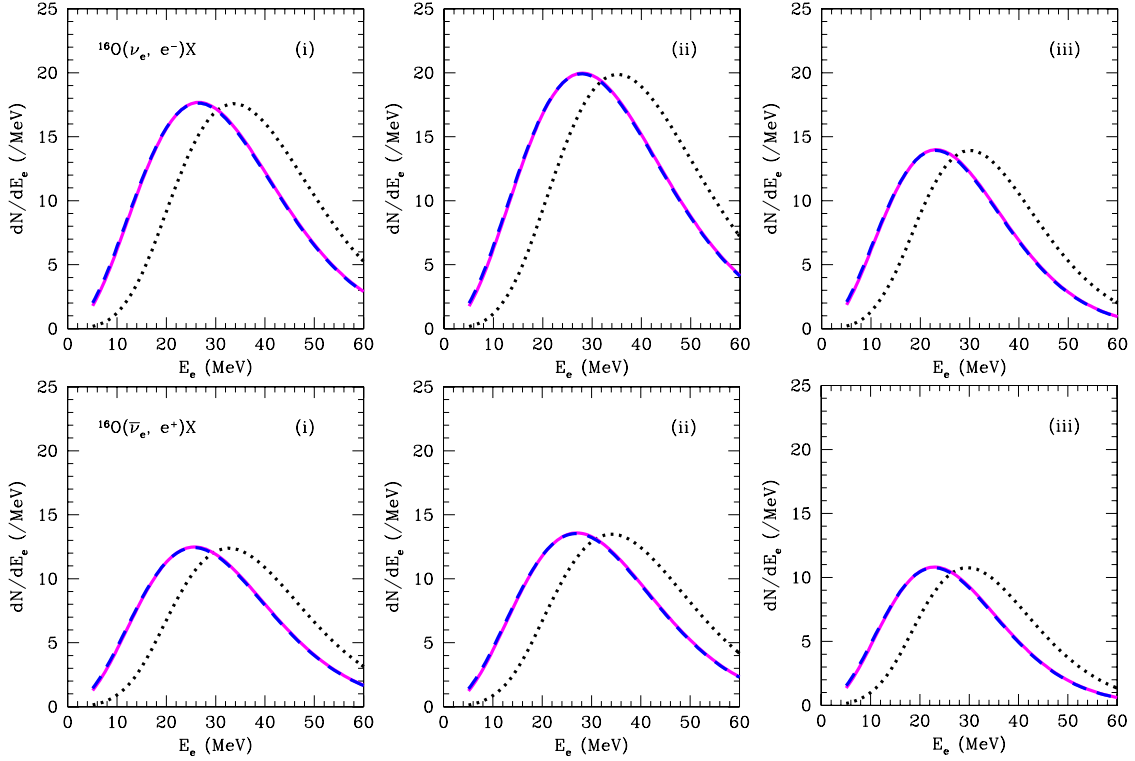}
\caption{Same as Fig.~\ref{ana10} but for $\langle E_{\nu_e}\rangle=\langle E_{\bar\nu_e}\rangle=20$~MeV.}
\label{ana20}
\end{figure}

The event spectra for the neutrino average energy $\langle E_{\nu_e}\rangle=\langle E_{\bar\nu_e}\rangle=20$~MeV are shown in Fig.~\ref{ana20}. We can again recognize that the ``four-group case'' is consistent with the ``42-state case'' but the ``single-energy case'' gives incorrect results and its spectra are about 7~MeV harder. For each condition, the total event number of the model with $\langle E_{\nu_i}\rangle=20$~MeV is larger than that with $\langle E_{\nu_i}\rangle=10$~MeV by about one order of magnitude. This large difference is attributed to the high threshold energy of neutrino-$^{16}$O charged-current reactions as already mentioned. The threshold energy of the $^{16}$O($\nu_e, e^-$)X reaction is higher than that of the $^{16}$O($\bar\nu_e, e^+$)X reaction. Therefore, for the models with $\langle E_{\nu_i}\rangle=10$~MeV, the $^{16}$O($\bar\nu_e, e^+$)X reaction has a larger event number than the $^{16}$O($\nu_e, e^-$)X reaction. Nevertheless, the opposite trend is seen for $\langle E_{\nu_i}\rangle=20$~MeV because the cross section of the $^{16}$O($\nu_e, e^-$)X reaction exceeds that of the $^{16}$O($\bar\nu_e, e^+$)X reaction for the high-energy regime, $E_\nu>35$~MeV. Note that the same spectrum (the total and average energies) is assumed for $\nu_e$ and $\bar\nu_e$ in these comparisons while it is inappropriate for realistic supernova neutrinos.

\subsection{Numerical case for supernova neutrino spectra}\label{spct2}
Here we examine the validity of our simplified approach by employing the numerical models of supernova neutrino spectra, which are taken from the Supernova Neutrino Database~\cite{self13a}. While several spectral models with different values of the progenitor mass $M$ and metallicity $Z$ are given in the Supernova Neutrino Database, here we adopt two sets: one is the model with $(M,Z)=(20M_\odot,0.02)$ and a shock revival time of 200~ms, which is chosen as an ordinary supernova neutrino model, and the other is the model with $(M,Z)=(30M_\odot,0.004)$, which is a model of neutrino emission from a black-hole-forming collapse. We consider the time-integrated spectra, whose average and total energies are listed in Table~\ref{nuspect}, and take into account neutrino oscillations as in Ref.~\cite{self15}. On the basis of Mikheyev-Smirnov-Wolfenstein effect resonances, the neutrino number spectra are written as
\begin{subequations}\label{nuspe}
\begin{equation}\label{nuspea}
\frac{d{\cal N}_{\nu_e}(E_\nu)}{dE_\nu} = P \frac{d{\cal N}^0_{\nu_e}(E_\nu)}{dE_\nu} + (1-P)\frac{d{\cal N}^0_{\nu_x}(E_\nu)}{dE_\nu},
\end{equation}
\begin{equation}\label{nuspeb}
\frac{d{\cal N}_{\bar\nu_e}(E_\nu)}{dE_\nu} = \bar P \frac{d{\cal N}^0_{\bar\nu_e}(E_\nu)}{dE_\nu} + (1-\bar P)\frac{d{\cal N}^0_{\nu_x}(E_\nu)}{dE_\nu},
\end{equation}
\end{subequations}
where $P$ and $\bar P$ are the survival probabilities of $\nu_e$ and $\bar\nu_e$, respectively~\cite{ds00}. Meanwhile, $d{\cal N}^0_{\nu_e}(E_\nu)/dE_\nu$, $d{\cal N}^0_{\bar \nu_e}(E_\nu)/dE_\nu$ and $d{\cal N}^0_{\nu_x}(E_\nu)/dE_\nu$ are neutrino number spectra before the oscillations adopted from the Supernova Neutrino Database, where $\nu_\mu$, $\nu_\tau$, $\bar\nu_\mu$ and $\bar\nu_\tau$ are collectively denoted as $\nu_x$. While the neutrino oscillation is caused by the mixture of flavor and mass eigenstates of neutrinos, there is still uncertainty concerning the mass ordering: normal ($m_1 < m_2 < m_3$) or inverted ($m_3 < m_1 < m_2$), where $m_1$, $m_2$ and $m_3$ are neutrino masses of the individual eigenstates. The survival probabilities depend on the mass hierarchy and we take $(P,\bar P)=(0,0.68)$ for a normal mass hierarchy and $(P,\bar P)=(0.32,0)$ for an inverted mass hierarchy. Then, the flux of $\nu_i$ is given as
\begin{equation}\label{nuflspe}
\frac{d{\cal F}_{\nu_i}(E_\nu)}{dE_\nu} = \frac{1}{4\pi d^2_\mathrm{SN}}\frac{d{\cal N}_{\nu_i}(E_\nu)}{dE_\nu}.
\end{equation}
Furthermore, we evaluate the event spectra not only for the neutrino-$^{16}$O charged-current channels but also for the electron-scattering and inverse $\beta$-decay channels. We use the cross sections in Ref.~\cite{totsuka94} for the electron scattering and in Ref.~\cite{strumia03} for the inverse $\beta$-decay.

\begin{table}[t]
\caption{Average energy, $\langle{E_{\nu_i}\rangle}$, and total energy, $E_{\nu_i,\mathrm{tot}}$, of the time-integrated $\nu_i$ spectrum, where $\nu_x=\nu_\mu$, $\nu_\tau$, $\bar\nu_\mu$, $\bar\nu_\tau$. The neutrino spectra of the ordinary supernova and the case of black hole formation are taken from the model with $(M,Z,t_{\rm revive})=(20M_\odot,0.02,200~{\rm ms})$ and the model with $(M,Z)=(30M_\odot,0.004)$ in Ref.~\cite{self13a}, respectively, where $M$ is the progenitor mass, $Z$ is the metallicity, and $t_{\rm revive}$ is the shock revival time.}
\label{nuspect}
\centering
\begin{tabular}{lcccccc}
\hline
 & $\langle{E_{\nu_e}\rangle}$ & $\langle{E_{\bar \nu_e}\rangle}$ & $\langle{E_{\nu_x}\rangle}$ & $E_{\nu_e,\mathrm{tot}}$ & $E_{\bar \nu_e,\mathrm{tot}}$ & $E_{\nu_x,\mathrm{tot}}$ \\
model & (MeV) & (MeV) & (MeV) & ($10^{52}$~erg) & ($10^{52}$~erg) & ($10^{52}$~erg) \\
\hline
 ordinary supernova & 9.32 & 11.1 & 11.9 & 3.30 & 2.82 & 3.27 \\
 black hole formation & 17.5 & 21.7 & 23.4 & 9.49 & 8.10 & 4.00 \\
\hline
\end{tabular}
\end{table}

\begin{figure}[t]
\centering\includegraphics[width=5.9in]{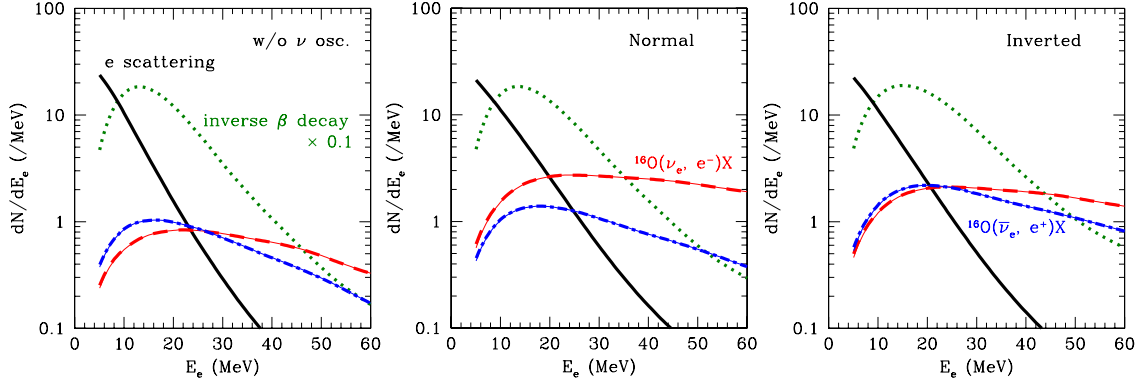}
\caption{Event spectra for the supernova neutrino model with $(M,Z)=(20M_\odot,0.02)$ and the shock revival time of 200~ms for the cases without the neutrino oscillation (left), with the normal hierarchy hypothesis (center) and with the inverted hierarchy hypothesis (right). Thick solid, dotted, dashed and dot-dashed lines correspond to electron-scattering, inverse $\beta$-decay, $^{16}$O($\nu_e, e^-$)X and $^{16}$O($\bar\nu_e, e^+$)X channels, respectively, where the spectra of the inverse $\beta$-decay reaction are multiplied by a factor of 0.1 and those of the $^{16}$O($\nu_e, e^-$)X and $^{16}$O($\bar\nu_e, e^+$)X reactions are for the ``four-group case''. Thin solid lines denote the neutrino-$^{16}$O charged-current spectra for the ``42-state case''.}
\label{intg2002}
\end{figure}

In Fig.~\ref{intg2002}, we show the event spectra for the ordinary supernova neutrino model. Comparing the ``42-state case'' and the ``four-group case'' for all models considered here, we find that the ``four-group case'' is sufficient to evaluate the $e^-$/$e^+$ spectra of the neutrino-$^{16}$O charged-current events. The expected event numbers are shown in Table~\ref{evnmb}. In the models with neutrino oscillation, flavor conversions from $\nu_\mu$ and $\nu_\tau$ to $\nu_e$ and from $\bar\nu_\mu$ and $\bar\nu_\tau$ to $\bar\nu_e$ occur. Then, the event numbers of the $^{16}$O($\nu_e, e^-$)X and $^{16}$O($\bar\nu_e, e^+$)X reactions increase because the average energy of $\nu_x$ is higher than those of $\nu_e$ and $\bar\nu_e$. In the case of the normal mass hierarchy, this effect is significant for $^{16}$O($\nu_e, e^-$)X reaction owing to the complete conversion. In contrast, for the $^{16}$O($\bar\nu_e, e^+$)X reaction, the event number in the case of the inverted mass hierarchy is larger than those of the other cases. These features are qualitatively consistent with the previous study~\cite{sk16}. Nevertheless, the present event spectra are softer because we take into account the excitation energies in our model.

\begin{table}[t]
\caption{Expected event numbers with a threshold energy of $E_e=5$~MeV for the models in Table~\ref{nuspect}.}
\label{evnmb}
\centering
\begin{tabular}{lrrrrrr}
\hline
 & \multicolumn{3}{c}{ordinary supernova} & \multicolumn{3}{c}{black hole formation} \\
reaction & no osc. & normal & inverted & no osc. & normal & inverted \\
\hline
$^{16}$O($\nu_e, e^-$)X &   41 &  178 &  134 &  2482 &  2352 &  2393 \\
$^{16}$O($\bar\nu_e, e^+$)X  &   36 &   58 &  103 &  1349 &  1255 &  1055 \\
electron scattering &  140 &  157 &  156 &   514 &   320 &   351 \\
inverse $\beta$-decay & 3199 & 3534 & 4242 & 17525 & 14879 &  9255 \\
total & 3416 & 3927 & 4635 & 21870 & 18806 & 13054 \\
\hline
\end{tabular}
\end{table}

We also show the event spectra of the electron-scattering and inverse $\beta$-decay channels in Fig.~\ref{intg2002}. Here the spectra of the inverse $\beta$-decay reaction are multiplied by a factor of 0.1, which corresponds to the unidentified events in SK-Gd under the assumption of 90\% neutron tagging efficiency as in Ref.~\cite{laha14a}. Since the electron scattering channel has a considerably different spectral shape from the other channels (Fig.~\ref{intg2002}) and a strong forward peak, the extraction of its signals is possible~\cite{laha14a}. However, even in SK-Gd, untagged inverse $\beta$-decay events may dominate and the spectral investigation of neutrino-$^{16}$O charged-current events is challenging for ordinary supernova neutrinos.

\begin{figure}[t]
\centering\includegraphics[width=5.9in]{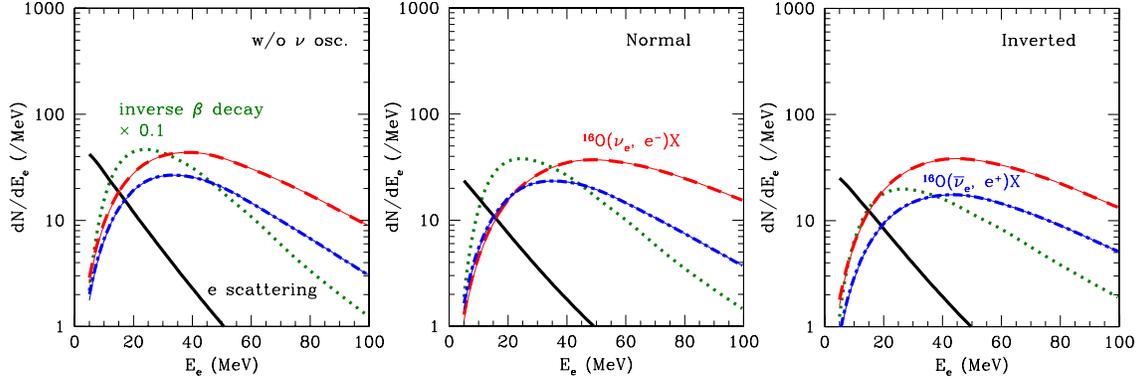}
\caption{Same as Fig.~\ref{intg2002} but for the model with $(M,Z)=(30M_\odot,0.004)$, which corresponds to a black-hole-forming collapse.}
\label{intg3010}
\end{figure}

For the case of a black-hole-forming collapse, the event spectra are shown in Fig.~\ref{intg3010}. It should be emphasized again that the neutrino-$^{16}$O charged-current spectra for the ``42-state case'' and ``four-group case'' coincide. The black hole formation model with an average neutrino energy of $\langle{E_{\nu_i}\rangle}\sim 20$~MeV has a larger event number than the ordinary supernova model with $\langle{E_{\nu_i}\rangle}\sim 10$~MeV by one order of magnitude, which is consistent with the results shown in Sect.~\ref{spct1} obtained from the analytic expressions for the neutrino spectra. In Fig.~\ref{intg3010}, the impact of neutrino oscillation is less than that for the ordinary supernova case (Fig.~\ref{intg2002}) for the following reason. In the black-hole-forming case, while the average energy of $\nu_x$ is again higher than those of $\nu_e$ and $\bar\nu_e$, the total emission energy of each $\nu_e$ and $\bar\nu_e$ is more than double that of $\nu_x$ (Table~\ref{nuspect}). Therefore, owing to neutrino oscillation, the neutrino spectrum becomes hard but the neutrino number decreases. As a result, these effects compensate each other and the impact of neutrino oscillation on the event spectrum is unremarkable as shown in Table~\ref{evnmb}.

The event spectra of the other channels are also shown in Fig.~\ref{intg3010}. Comparing the black hole formation model with the ordinary supernova model, we can see that the event number of the inverse $\beta$-decay reaction is increased but the enhancement ratio is less than that of the neutrino-$^{16}$O charged-current reactions. Meanwhile, if the events of the inverse $\beta$-decay reaction are identified with an efficiency of 90\% by neutron tagging in SK-Gd, the event number of neutrino-$^{16}$O charged-current reactions and the number of untagged inverse $\beta$-decay events become comparable. Furthermore, using the spectrum of the identified inverse $\beta$-decay events, we can statistically subtract the contribution of the untagged inverse $\beta$-decay events~\cite{laha14a} to isolate the signal of the neutrino-$^{16}$O charged-current reactions. In addition, owing to the high event energies, the contamination from extra $\gamma$ rays by decaying daughter nuclei would be resolved while the accuracy of the event reconstruction is desirable to be investigated for future work. Therefore, the neutrino-$^{16}$O charged-current channels are clearly worth investigation for the future detection of a neutrino burst with high average energies, as in the case of a black-hole-forming collapse.

\section{Conclusion}\label{conc}
In this paper, the neutrino-$^{16}$O charged-current reactions in a large water Cherenkov detector are studied as a detection channel for supernova neutrinos. To evaluate the event spectra, we consider the cross section for each excitation energy and the energy distribution of the recoil $e^-$/$e^+$, which were not dealt with in previous studies. For this purpose, we employ shell model calculations based on the SFO-tls Hamiltonian. As a result, we find that the event spectra obtained in this study are softer than those in previous studies because a considerable fraction of neutrino-$^{16}$O charged-current reactions occur with an excitation energy clearly higher than the threshold energy. Furthermore, while 42 states are selected in the shell model calculations, we propose a simplified approach to quickly obtain the spectra of the recoil $e^-$/$e^+$. The 42 states are divided into four energy groups in this approach and we confirm that the simplified approach is valid in the modeling of the event spectra employing both analytic expressions and numerical models for the supernova neutrino spectra. The partial cross sections of the four energy groups are provided both in table form (Table~\ref{cslist}) and as fitting formula (\ref{csfit}), which we believe will be useful for future studies. Because of their high energy threshold, the neutrino-$^{16}$O charged-current reactions are expected to be useful for diagnosing the average energy of neutrinos in a future SK-Gd experiment, where their events will be distinguished from those of the inverse $\beta$-decay reaction owing to the neutron tagging by Gd.

\section*{Acknowledgment}
The authors are grateful to K. Inoue and K. Okumura for valuable comments. This work was partially supported by JSPS KAKENHI Grant Numbers JP26104001, JP26104006, JP15K05090 and JP17H05203.


%

%

\end{document}